\documentclass[twocolumn,showpacs,preprintnumbers,amsmath,amssymb,superscriptaddress,prb]{revtex4}

\usepackage{graphicx}
\usepackage{dcolumn}
\usepackage{bm}
\usepackage{color}
\usepackage{ulem}

\begin{document}

\title{Combining photoemission and optical spectroscopies for reliable valence determination in YbS and Yb metal}

\author{M.~Matsunami}
 \altaffiliation[Electronic address: ]{matunami@spring8.or.jp}
 \affiliation{Institute for Solid State Physics, The University of Tokyo, Kashiwa, Chiba 277-8581, Japan}
 \affiliation{RIKEN SPring-8 Center, Sayo-cho, Sayo-gun, Hyogo 679-5148, Japan}

\author{A.~Chainani}
\author{M.~Taguchi}
\author{R.~Eguchi}
\author{Y.~Ishida}
\author{Y.~Takata}
 \affiliation{RIKEN SPring-8 Center, Sayo-cho, Sayo-gun, Hyogo 679-5148, Japan}

\author{H.~Okamura}
\author{T.~Nanba}
\affiliation{Department of Physics, Graduate School of Science, Kobe University, Kobe 657-8501, Japan}

\author{M.~Yabashi}
 \affiliation{RIKEN SPring-8 Center, Sayo-cho, Sayo-gun, Hyogo 679-5148, Japan}
 \affiliation{JASRI/SPring-8, Sayo-cho, Sayo-gun, Hyogo 679-5198, Japan}

\author{K.~Tamasaku}
\author{Y.~Nishino}
 \affiliation{RIKEN SPring-8 Center, Sayo-cho, Sayo-gun, Hyogo 679-5148, Japan}

\author{T.~Ishikawa}
 \affiliation{RIKEN SPring-8 Center, Sayo-cho, Sayo-gun, Hyogo 679-5148, Japan}
 \affiliation{JASRI/SPring-8, Sayo-cho, Sayo-gun, Hyogo 679-5198, Japan}

\author{Y.~Senba}
\author{H.~Ohashi}
 \affiliation{JASRI/SPring-8, Sayo-cho, Sayo-gun, Hyogo 679-5198, Japan}

\author{N.~Tsujii}
\affiliation{National Institute for Materials Science, 1-2-1 Sengen, Tsukuba, Ibaraki 305-0047, Japan}

\author{A.~Ochiai}
\affiliation{Center for Low Temperature Science, Tohoku University, Aoba-ku, Sendai 980-8578, Japan}

\author{S.~Shin}
 \affiliation{Institute for Solid State Physics, The University of Tokyo, Kashiwa, Chiba 277-8581, Japan}
 \affiliation{RIKEN SPring-8 Center, Sayo-cho, Sayo-gun, Hyogo 679-5148, Japan}

\date{\today}

\begin{abstract}
Hard x-ray photoemission and optical spectroscopies have been performed on YbS and Yb metal to determine the precise $f$-electron occupation. 
A comparison of the photoemission spectra with the energy loss functions in bulk and surface, obtained from optical reflectivity, enables us to distinguish between the energy loss satellite of Yb$^{2+}$ peak and Yb$^{3+}$ multiplet. 
The results clearly indicate a purely divalent Yb state except for the surface of YbS. 
We demonstrate that the present method is highly reliable in identifying the electronic structure and the mean valence in $f$-electron systems. 

\end{abstract}

\pacs{71.27.+a, 79.60.-i, 75.30.Mb}

\maketitle

In the field of correlated $f$-electron physics, the concept of mixed valence has been crucial in terms of the $f$ electrons at the boundary between itinerant and localized states. \cite{Mixed_Valence} 
This is closely related to the heavy-fermion or quantum critical behavior in $f$-electron systems. 
It is therefore important to correctly determine the mean valence of the rare-earth ion, since the mean valence is a direct measure of the degree of $f$-electron localization. 
High energy spectroscopy such as photoemission spectroscopy (PES) and x-ray absorption spectroscopy (XAS) has been a fundamental probe to directly observe the electronic structure including mean valence determination. 
Nonetheless, the reported valence values have been often inconsistent, particularly with respect to basic physical properties, as discussed in the following.

YbS and Yb metal have been considered as prototypical divalent Yb systems at ambient pressure. \cite{Yb_Theory} 
However, a small amount of trivalent signal has been observed in some studies. 
For YbS, Yb $L_{\rm III}$-edge XAS reported +2.08 as the mean valence for Yb. \cite{YbS_Syassen} 
Moreover, recent XAS and Yb $L_{\rm \alpha 1}$ resonant inelastic x-ray scattering (RIXS) on powder samples reported a higher valence (+2.35$\pm$0.04). \cite{YbS_RIXS} 
Also for Yb metal, the Yb valence has been uncertain between +2 and +2.1 for some XAS and RIXS results. \cite{Yb_Syassen,Yb_XANES,Yb_RIXS} 
In addition, the valence-band PES using a photon energy $h\nu$=40.8\,eV for Yb metal concluded the $f$-electron count $n_f$=13.98 from the analysis of single impurity Anderson model. \cite{Yb_VB_PES_Patthey} 
It is important to clarify whether YbS and Yb metal have intrinsically mixed-valence ground state or not. 
In particular, YbS has been identified as an ionic and divalent insulator, Yb$^{2+}$S$^{2-}$, with an energy gap of $\sim$1.2\,eV between fully occupied Yb\,4$f$ states and unoccupied Yb\,5$d$ states. \cite{Optical_YbS_by_Syassen} 
Only under high pressure above $\sim$10\,GPa, the mixed valence state is realized, coupled with the insulator-to-metal transition. \cite{Optical_YbS_by_Syassen,Optical_YbS_by_us} 
Thus, it seems contradictory that XAS and RIXS indicate mixed valence in YbS coexisting with the insulating ground state. 
The estimation for the stability of divalent state in YbS and Yb metal gives fundamental information to understand the mixed-valence state under high pressure, and more generally, provides important inputs for modeling the valence fluctuation and transition in $f$-electron systems.

Compared to XAS and RIXS, the rare-earth 3$d$ core-level PES is the suitable probe to correctly determine the mean valence in terms of a well-separated divalent and trivalent state. 
However, Yb\,3$d$ core levels, located at the binding energy of 1500-1600\,eV, could not be excited by a conventional soft x-ray (SX) source (e.g. Al-$K\alpha$ radiation, $h\nu$=1.49\,keV), while there are a few studies using Si-$K\alpha$ radiation ($h\nu$=1.74\,keV). \cite{YbP_SiKa1,Yb_SiKa1,Yb_SiKa2,Yb3d_PES_Patthey} 
Recent advances in hard x-ray (HX) PES make it possible to observe not only the valence band but also the Yb\,3$d$ core levels under bulk sensitive condition, based on the increased probing depth. \cite{YbInCu4_HXPES_Sato,YbAl3_HXPES_Suga}

In this paper, we report the HX-PES of Yb\,3$d$ core levels and valence band for YbS and Yb metal. 
The measured HX-PES spectra are compared with the energy loss functions in the bulk and surface, measured with optical spectroscopy, to analyze the electronic structure and mean valence in YbS and Yb metal. 
For comparison, HX-PES spectra of a typical mixed-valence compound YbCu$_2$Si$_2$ are also measured. 
The results clearly indicate the purely divalent Yb state and strictly exclude the mixed-valence ground state in bulk YbS and Yb metal.

Single crystals for YbS and YbCu$_2$Si$_2$ were grown by a Bridgman method and an indium-flux method, respectively. 
Their clean surfaces were obtained by cleaving $in$ $situ$. 
Yb metal (99.99\,\% purity) was prepared as films by $in$ $situ$ evaporation under ultra-high vacuum condition. 
HX-PES experiments were carried out at undulator beamline BL29XUL in SPring-8, using $h\nu$=7.94\,keV and a hemispherical electron analyzer Scienta R4000-10kV. \cite{BL29XUL} 
For the valence band of Yb metal, SX-PES was also performed at undulator beamline BL17SU, using $h\nu$=1.2\,keV and Scienta SES-2002. \cite{BL17SU} 
The total energy resolution was set to $\sim$200\,meV for both HX- and SX-PES. 
To obtain the energy loss functions of YbS and Yb metal, their optical reflectivity spectra were measured between far-infrared and vacuum ultraviolet region. 
The obtained spectra were consistent with the previous data. \cite{Optical_YbS_by_Wachter,Optical_BaSrEuYb} 
The Kramers-Kronig analysis was used to calculate the complex dielectric function $\varepsilon$ from the reflectivity. 
The obtained $\varepsilon$ was then used to calculate the bulk and surface energy loss functions, defined as Im(1/$\varepsilon$) and Im[1/($\varepsilon$+1)], respectively. \cite{Surface_Loss}

\begin{figure}[t]
\begin{center}
\includegraphics[width=0.4\textwidth]{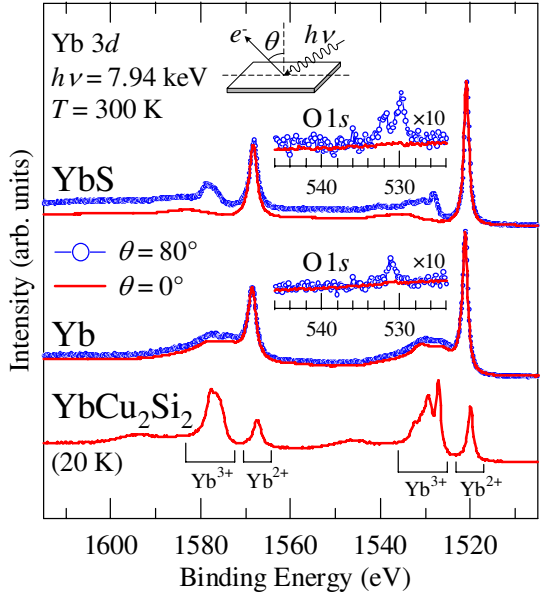}
\caption{
(Color online) 
The Yb\,3$d$ core-level spectra for YbS and Yb metal recorded at the emission angle $\theta$\,=\,0$^\circ$ and 80$^\circ$, in comparison with that of YbCu$_2$Si$_2$ ($\theta$\,=\,0$^\circ$). 
Each inset shows the $\theta$ dependence of O\,1$s$ core-level spectra in the magnified scale. 
The definition of $\theta$ is also shown in the upper inset. 
} 
\end{center}
\end{figure}

Figure~1 shows the Yb\,3$d$ core-level spectra for YbS and Yb metal recorded at the emission angle $\theta$\,=\,0$^\circ$ and 80$^\circ$, in comparison with that of YbCu$_2$Si$_2$ at $\theta$\,=\,0$^\circ$. 
Since the $\theta$ is defined as shown in the upper inset of Fig.~1, $\theta$\,=\,0$^\circ$ and 80$^\circ$ are bulk and surface sensitive configurations, respectively. 
These spectra are separated into 3$d_{5/2}$ region at 1515-1540\,eV and 3$d_{3/2}$ region at 1560-1585\,eV by the spin-orbit splitting. 
In addition, for YbCu$_2$Si$_2$, both final states are split into 3$d^9$4$f^{14}$ (Yb$^{2+}$) line (at $\sim$1520 and $\sim$1568\,eV) and 3$d^9$4$f^{13}$ (Yb$^{3+}$) multiplet (at $\sim$1525-1538 and $\sim$1572-1583\,eV) due to the valence fluctuation. 
From the ratio of their integrated intensities, the Yb valence for YbCu$_2$Si$_2$ can be estimated as +2.8, which agrees well with its bulk value. \cite{RIXS_vs_HXPES} 
As is clear from the spectral shape, the Yb$^{3+}$ multiplet feature also appears in the 80$^\circ$ spectrum for YbS. 
In contrast, the spectra at 0$^\circ$ for YbS and both angles for Yb metal are composed of the Yb$^{2+}$ peak and its broad satellite feature as discussed later. 
The $\theta$ dependence of O\,1$s$ core-level spectra is also shown in the insets of Fig.~1. 
The O\,1$s$ signal is completely suppressed or negligibly weak at 0$^\circ$ for both YbS and Yb metal. 
On the other hand, for 80$^\circ$ spectra, the weak O\,1$s$ signal can be observed as two peaks and one peak for YbS and Yb metal, respectively. 
By analogy with the oxidation behavior of Cr metal \cite{Cr_metal} or Pr metal, \cite{Pr_metal} the 532-eV peak in YbS can be identified as an indicator of the growth for Yb$_2$O$_3$ due to the surface oxidation, resulting in the trivalent signal for 80$^\circ$ spectrum of Yb\,3$d$ core levels. 
On the other hand, the 530-eV peak in YbS and the 531-eV peak in Yb metal may be due to the isolated oxygen physisorption and not related with the bulk Yb valence.

\begin{figure}[t]
\begin{center}
\includegraphics[width=0.47\textwidth]{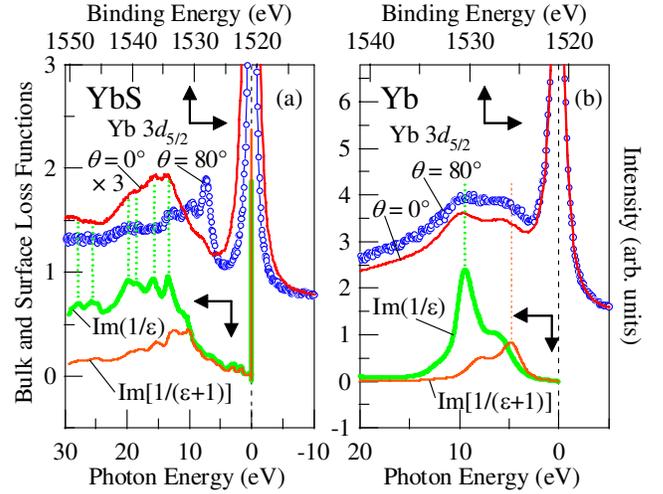}
\caption{
(Color online) 
The satellite structures in Yb\,3$d_{5/2}$ core-level spectra for YbS (a) and Yb metal (b), in comparison with bulk loss function Im(1/$\varepsilon$) and surface loss function Im[1/($\varepsilon$+1)] where $\varepsilon$ is the complex dielectric function obtained from the measured optical reflectivity. 
The dotted lines indicate main peak positions in loss functions. 
} 
\end{center}
\end{figure}

Figure~2 compares the satellite features in Yb\,3$d_{5/2}$ core-level spectra of YbS and Yb metal with their bulk and surface loss functions, where the loss functions are plotted relative to the Yb$^{2+}$\,3$d_{5/2}$ peak energy. 
For YbS, the satellite feature in the 0$^\circ$ spectrum shows good correspondence with the peaks in the bulk loss function. 
Hence, the satellite is attributed to the energy loss caused by interband transitions, which give main contributions to the optical response in this photon-energy range. 
On the other hand, the main feature in 80$^\circ$ spectrum shows no such correspondence with the energy loss functions, thus indicating that the feature in the 80$^\circ$ spectrum is mainly due to Yb$^{3+}$\ multiplet. 
For Yb metal, both loss functions are mainly composed of two features, which are due to the plasmon excitations at 8-10\,eV and the interband transitions at 5-7\,eV. \cite{Optical_BaSrEuYb} 
The satellite features in 0$^\circ$ spectrum can be well reproduced by combination of the bulk and surface loss functions. 
The surface loss weight in the 80$^\circ$ spectrum is slightly enhanced compared with that in the 0$^\circ$ one, which is reasonable. 
In addition, the spectral shapes of the satellites at both 0$^\circ$ and 80$^\circ$ are very different from that of the Yb$^{3+}$ multiplet. 
Hence, the $\theta$ dependence of Yb\,3$d$ core-level spectra for Yb metal should result from the weight variation between bulk and surface loss components associated with different probing depths at 0$^\circ$ and 80$^\circ$. 
This interpretation is consistent with the previous SX-PES result for Yb metal, \cite{Yb3d_PES_Patthey} in which the surface loss dominant portion (at $\sim$5\,eV) of the satellite is stronger than the bulk loss dominant one (at $\sim$10\,eV), because of the small probing depth compared with the present HX-PES.

\begin{figure}[t]
\begin{center}
\includegraphics[width=0.45\textwidth]{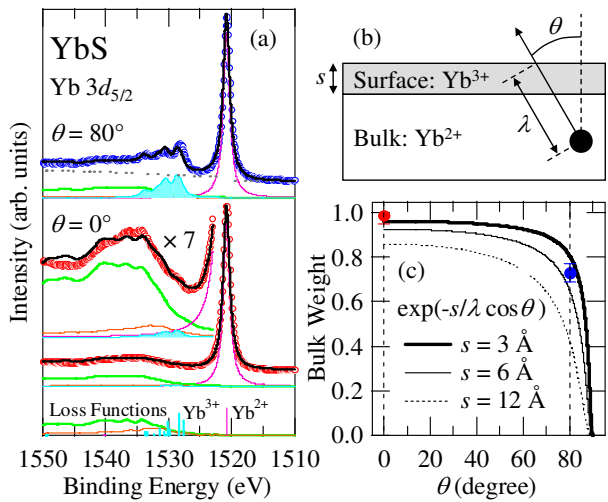}
\caption{(Color online) 
(a) Fitting of the emission-angle $\theta$ dependent Yb\,3$d_{5/2}$ core-level spectra for YbS by using lifetime broadened Yb$^{2+}$ line, Yb$^{3+}$ multiplet, and loss functions. 
The integral background is used only for $\theta$\,=\,80$^\circ$ as shown by the dotted line. 
(b) The simple model for bulk and surface Yb valence in YbS, and the photoemission configuration. 
(c) The $\theta$ dependence of bulk weight in photoemission spectra of YbS. 
} 
\end{center}
\end{figure}

\begin{figure}[t]
\begin{center}
\includegraphics[width=0.4\textwidth]{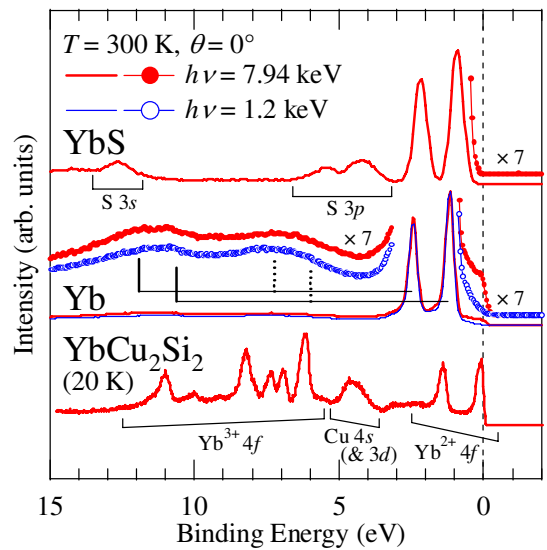}
\caption{
(Color online) The valence-band spectra for YbS, Yb metal, and YbCu$_2$Si$_2$. 
For Yb metal, the soft x-ray data is also plotted. 
The vertical bars represent the main peak positions of bulk loss function (solid bar) and surface loss function (dotted bar), relative to Yb$^{2+}$\,4$f_{5/2}$ and 4$f_{7/2}$ peak. 
} 
\end{center}
\end{figure}

In order to estimate the bulk and surface Yb valence for YbS, the Yb\,3$d$ core-level spectra at 0$^\circ$ and 80$^\circ$ are fitted by using lifetime broadened Yb$^{2+}$ line, Yb$^{3+}$ multiplet, \cite{atomic_calc} and (bulk and surface) loss functions, as shown in Fig.~3(a). 
A very small Yb$^{3+}$ weight is required to reproduce the spectral feature around 1528\,eV in 0$^\circ$ spectrum. 
This suggests that 0$^\circ$ spectrum also contains tiny amount of surface signal. 
From this analysis, the mean valence for Yb is evaluated as +2.01 and +2.27 at 0$^\circ$ and 80$^\circ$, respectively. 
To simplify the analysis, we assume Yb valence to be trivalent in surface and divalent in bulk, as shown in Fig.~3(b). 
In this case, the divalent weight of the estimated Yb mean valence is equivalent to the bulk weight in the spectra. 
Generally, the bulk weight in PES spectra is expressed as exp[-$s$/($\lambda$cos$\theta$)], where $s$ and $\lambda$ denote the thickness of the surface layer and the inelastic mean free path of photoelectron, respectively. 
Figure~3(c) shows the $\theta$ dependence of bulk weight calculated by using $\lambda$\,=\,80\,\AA~(Ref.~\onlinecite{Probing_Depth}) and $s$\,=\,3, 6, and 12\,\AA. 
The estimated bulk weight for YbS is located between $s$\,=\,3 and 6\,\AA. 
Taking into account the lattice constant of YbS (=\,5.68\,\AA), \cite{YbS_Lattice} it is found that only the top two layers are oxidized and turn into trivalent state. 
In addition, the trivalent signal in Yb\,3$d$ core-level spectra gradually increases with time even under ultra-high vacuum condition (not shown), representing an increase of the oxidized layer. 
These results suggest that the mixed valence in YbS observed by XAS and RIXS \cite{YbS_RIXS} may be resulting from the surface oxidation of powder samples.

A purely divalent Yb state in YbS and Yb metal is also evident from their valence-band spectra shown in Fig.~4, which also contains data for YbCu$_2$Si$_2$. 
The Yb$^{2+}$-4$f$ doublet ($J$=5/2 and 7/2) corresponding to 4$f^{13}$ final state is positioned at or slightly below Fermi level ($E_{\rm F}$). 
The peaks for YbS are symmetrically broadened compared to those for Yb metal, as is generally observed in ionic crystals. \cite{phonon_broadening} 
The Yb$^{3+}$-4$f$ multiplet corresponding to 4$f^{12}$ final state can be observed at 5-13\,eV for YbCu$_2$Si$_2$, but not for YbS and Yb metal. 
For Yb metal, the two weak and broad features centered at 6-7 and 11-12~eV are observed. 
It is seen that the positions of these features well correspond to those of the main peaks in the bulk and surface loss functions relative to Yb$^{2+}$\,4$f_{5/2}$ and 4$f_{7/2}$ peaks, as indicated by vertical bars in Fig.~4. 
Thus, these features can be also attributed to the bulk and surface loss satellites as in the case of Yb\,3$d$ core-level PES [Fig.~2(b)]. 
For YbS, the energy loss satellites may be too small to be observed in the valence-band spectrum.

The density of states (DOS) at $E_{\rm F}$ gives further piece of evidence for a purely divalent Yb state. 
In the case of mixed-valence systems, the Yb$^{2+}$-4$f_{7/2}$ peak is located just at $E_{\rm F}$ and is identified as a Kondo resonance peak, as clearly seen in YbCu$_2$Si$_2$ (Fig.~4). 
On the other hand, for YbS and Yb metal, the 4$f_{7/2}$ peak is located away from $E_{\rm F}$. 
For YbS, the absence of a Fermi edge is consistent with fully occupied 4$f$ picture in the insulating ground state. 
For Yb metal, the Fermi edge is clearly observed in the spectrum at $h\nu$=7.94\,keV. 
Previously, a Fermi edge was also reported in the spectrum at $h\nu$=40.8\,eV and was attributed to a cutting off of the tail of 4$f_{7/2}$ peak. \cite{Yb_VB_PES_Patthey} 
This assignment inevitably required a finite unoccupied 4$f$ DOS, which led to a weakly mixed-valence state with $n_f$=13.98. \cite{Yb_VB_PES_Patthey} 
However, the spectral intensity around $E_{\rm F}$ is significantly suppressed at $h\nu$=1.2\,keV as shown in Fig.~4. 
According to the photoionization cross sections ($\sigma$), \cite{Cross_Section} the ratio $\sigma$(Yb\,6$s$)/$\sigma$(Yb\,4$f$) at $h\nu$=40.8\,eV or 7.94\,keV is much greater than that at $h\nu$=1.2\,keV. 
Hence, the DOS at $E_{\rm F}$ in Yb metal should be attributed to the 6$s$ electrons rather than the tail of 4$f_{7/2}$ peak. 
In fact, HX-PES is well known as an $s$-electron sensitive probe. \cite{HX-PES_Ag} 
Consequently, the present results clearly indicate that the 4$f$ states in Yb metal are also fully occupied.

Finally, we comment on the stability of divalent Yb state. 
According to an $ab$ $initio$ band structure study, the stability of divalent Yb state for YbS and Yb metal depends on the treatment of 4$f$ electrons; the divalent state in YbS is more (less) stable than that in Yb metal within the core-like (band-like) treatment. \cite{Yb_Theory} 
In the valence-band spectra in Fig.~4, the position of Yb$^{2+}$-4$f$ final state is closer to $E_{\rm F}$ in YbS than in Yb metal. 
This observation confirms the validity of the band-like treatment of 4$f$ electrons in Yb-based systems. 
Simultaneously, it can be verified that 4$f$ electrons provide no contribution to the DOS at $E_{\rm F}$ for Yb metal.

In summary, we have performed the HX-PES for YbS and Yb metal. 
The bulk Yb\,3$d$ core-level spectra are composed of Yb$^{2+}$ peak and its satellite feature. 
The satellite is well explained by the energy loss functions in bulk and surface, obtained from optical spectroscopy. 
For YbS, the Yb$^{3+}$ component due to the oxidation is observed only in the surface. 
The valence-band spectra of Yb metal in comparison with the SX-PES data reveal that the DOS at $E_{\rm F}$ is dominated by mainly 6$s$ electrons excluding the 4$f$ electrons. 
The results unambiguously elucidate the purely divalent Yb state in bulk YbS and Yb metal.



\end{document}